\documentclass[a4paper,11pt]{article}
\usepackage[T1]{fontenc}
\usepackage{jinstpub} 
\usepackage{lineno}
\usepackage{listings}
\usepackage{subcaption}
\usepackage{enumerate}
\usepackage{wrapfig}
\usepackage{tabularx}
\usepackage[english]{babel}
\usepackage{siunitx}
\usepackage{graphicx} 
\usepackage{amsmath} 
\usepackage{mathtools}
\usepackage{bbold}
\usepackage{hyperref}
\usepackage{titlesec}
\usepackage{setspace} 

\title{DEDALO: Device for Enhanced Dust Analyses with Light Obscuration sensors}

\author[a, 1]{Luca Teruzzi,\note{Corresponding author.}}
\author[b]{, Llorenç Cremonesi}
\author[a]{, and Marco A.C. Potenza}

\affiliation[a]{Department of Physics, University of Milan, Via Celoria 16, 20133 Milan, Italy}
\affiliation[b]{Department of Earth and Environmental Sciences, University of Milan-Bicocca, Piazza della Scienza 1, 20126 Milan, Italy}

\emailAdd{luca.teruzzi@unimi.it}

\abstract{Instruments based on light obscuration sensors are widely used for measuring the size distribution of insoluble sub-visible particles in liquid suspensions, being fast and suitable for in situ and real-time measurements. Such instruments are typically calibrated by means of reference polystyrene spherical particles with a specific refractive index, which unavoidably leads to systematic errors when determining the size of particles of different materials. In this paper, we propose a reliable and consistent method to overcome this limitation by setting the refractive index value according to the sample, thus achieving an improved particle size distribution (PSD) measurement. 
An ad hoc, ready-to-use, open source code with a graphical interface able to drive an in-line instrument and obtain a real-time correction to the PSD has been developed. The method has been extensively validated with several oil emulsions characterized by different refractive index values and the results have been compared with an independent optical method. As an example of application, we have adopted this approach for the analysis of dust suspended in meltwater of an ice core from a glacier in the Aosta Valley (Italy). We believe that our approach will strongly improve the accuracy in characterizing liquid suspensions and reduce discrepancies between data obtained with different methods. The code has been made publicly available at: \url{https://instrumentaloptics.fisica.unimi.it/dedalo/} and on the GitHub page of the corresponding author (\url{https://github.com/LucaTeruzzi/DEDALO}).}


\begin{document}
\maketitle
\flushbottom


\section{Introduction}
\label{sec:intro}

Light obscuration is one of the most reliable methods for measuring the size distribution (PSD) and number concentration of sub-visible particles suspended in a liquid \cite{USP}. It is also the European Union reference for sizing, according to the ISO21501-3 standard \cite{ISO}. The liquid suspension is driven through a small cell transilluminated by a laser beam. When a microparticle intercepts the beam, a photodiode detects the corresponding attenuation, thus measuring the optical extinction cross-section, $C_\mathrm{ext}$. According to general light scattering laws, $C_\mathrm{ext}$ is mainly affected by particle size \cite{VdH, BH}, therefore, it is retrieved with relative ease if the refractive index is established first. Instruments based on this working principle are often called Light Obscuration (LO) sensors. They are able to characterize a wide range of particles, as small as \SI{0.2}{\micro\meter} in diameter. Additionally, they are fast, easy to use, and can be used in situ or with minimal sample preparation. Therefore, they are suitable for real-time monitoring and process control applications \cite{LO_applcation1, LO_applcation2}. They provide results within seconds, which is crucial for industrial use, including pharmaceuticals and electronics applications. 

The main drawback of instruments providing such mono-parametric measurements is that the refractive index cannot usually be inferred from the light scattering data. In fact, it is customary to assume a one-fits-all value, $n_0$, which might differ from that of the actual particles under study; no absorption is considered. The standard calibration procedure is usually based on light scattering by polystyrene spheres: it is assumed that the size of a particle is that of the polystyrene sphere that causes the measured attenuation of the laser beam. However, the refractive index $n$ of particles that populate real samples can appreciably differ from the reference value $n_0$ and can depend on temperature \cite{refindex_temp1, refindex_temp3} and pressure \cite{refindex_temp2}. Generally speaking, particles with a larger refractive index extinguish more light than particles with a lower refractive index of the same size. This implies that the size of particles characterized by $n<n_0$ will be underestimated and vice versa \cite{ref_index_effect}. Moreover, the larger the refractive index deviation from the calibration standard, the less accurate the PSD results are \cite{LO_solution_prop}. A larger difference in the refractive index from that of the surrounding medium accentuates this effect \cite{LO_solution_prop}, some examples being air bubbles in water or oil droplets in water.

Other particle sizing techniques such as dynamic light scattering (DLS), laser diffraction, or Small Angle Light Scattering (LALS) may be used as complementary methods to provide additional information on particle size distribution \cite{LO_comparison, LO_bias_correction}, at the cost of reducing the effectiveness in terms of data throughput. In some cases, the refractive index can be measured or inferred independently \cite{LO_modified}. 

In this paper, we present a ready-to-use software to overcome the limitation of having to fix the refractive index a priori. An open-source Python-based GUI code allows the user to operate an in-line LO instrument and retrieve in real-time the correct PSDs. The algorithm operates on LO data with the refractive index as a free parameter to be set according to the sample composition. The code is named DEDALO (\textit{Device for Enhanced Dust Analyses with Light Obscuration sensors}). 

We validate the method with a range of laboratory-prepared oil emulsions characterized by different refractive index values, ranging from 1.46 to 1.64. Samples have then been analyzed through a Continuous Flow Analysis (CFA) system with a commercial LO device (Abakus particle counter, Klotz GmbH, Germany). An independent instrument (Single Particle Extinction and Scattering method, SPES) installed downstream the Abakus provides both the PSDs and the refractive index, which in principle is known for the samples under study \cite{spes_1, spes_2, spes_4}. We point out that all the tested samples are characterized by a real refractive index, thus absorption is negligible in our work.

This study builds on a collaboration between the Instrumental Optics Laboratory of the University of Milan and the EuroCold laboratory of the glaciology and palaeoclimate group at the University of Milan-Bicocca, where all these methods are adopted for characterizing micron-sized dust from ice cores. After validating our code extensively, we performed high-resolution and high throughput measurements of mineral dust in meltwater from cryosphere samples collected on an Alpine glacier currently under study by our group. Beyond the specific application discussed in this work, we believe that the applications of DEDALO could be far more general, ranging from pharmaceutics to water quality assessments and clinical assays. 


\section{Materials and methods}
\label{sec:materials_and_methods}

The samples are oil-in-water emulsions prepared by adding \SI{1}{\micro\liter} of oil with calibrated refractive index (Cargille laboratories \cite{cargille}) to \SI{50}{\milli\liter} of MilliQ water (resistivity \SI{18.2}{\mega\ohm\cdot\centi\meter}, total organic carbon 2 ppb), then shaking the sample. We chose oils with a refractive index ranging from 1.46 to 1.64 ($\pm$ 0.002 uncertainty, standard measure at \SI{589.3}{\nano\meter} and \SI{25}{\celsius}). Specifically, we prepared eleven samples with a refractive index of 1.46, 1.47, 1.48, 1.50, 1.51, 1.52, 1.53, 1.56, 1.58, and 1.64. The emulsions we obtained have similar PSDs with modes around \SI{1}{\micro\meter}, reasonably described by log-normal distributions (see below in Section \ref{sec:data}). We also prepared two additional samples that contain two independent populations each, by mixing two different immiscible oils in pure water (refractive indices 1.50 and 1.52, 1.50 and 1.56). They have been used as a stress test for the algorithm and verify its ability to operate with mixed emulsions, where at least one of the species (actually both of them) deviates from the ideal refractive index of calibration particles. This is of particular interest in real-world situations, where the refractive index is often unknown and the sample may include a mixture of particles with different refractive indices. Table \ref{tab:emulsions} summarizes the numeric particle concentrations and the corresponding mode for each laboratory-prepared sample.

\begin{table}[htbp]
\renewcommand{\arraystretch}{1.1}
\centering
\caption{Nominal refractive index ($\lambda=\SI{589.3}{\nano\meter}$, \SI{25}{\celsius}), numeric particle concentrations and PSDs modes measured for all the samples prepared to validate DEDALO effectiveness.\label{tab:emulsions}}
\smallskip
\begin{tabular}{c|c|c}
\hline\hline
\hspace{0.2cm} \textbf{Refractive index} \hspace{0.2cm} & \hspace{0.2cm} \textbf{Sample concentration [ptc/mL]} \hspace{0.2cm} & \hspace{0.2cm} \textbf{Mode [\si{\micro\meter}]} \hspace{0.2cm} \\
\hline\hline
1.46 & 5.4$\cdot$10$^5$ & 1.5 \\
1.47 & 1.2$\cdot$10$^6$ & 1.4 \\
1.48 & 2.1$\cdot$10$^5$ & 1.4 \\
1.50 & 1.2$\cdot$10$^6$ & 1.2 \\
1.51 & 1.6$\cdot$10$^6$ & 1.2 \\
1.52 & 1.6$\cdot$10$^6$ & 1.1 \\
1.53 & 2.2$\cdot$10$^6$ & 1.0 \\
1.56 & 1.1$\cdot$10$^6$ & 0.9 \\
1.58 & 2.7$\cdot$10$^6$ & 0.9 \\
1.64 & 1.1$\cdot$10$^6$ & 0.8 \\
\hline
1.50 and 1.52 & 1.4$\cdot$10$^6$ & 1.1 \\
1.50 and 1.56 & 1.5$\cdot$10$^6$ & 0.9 \\
\hline\hline
\end{tabular}
\end{table}

\begin{figure}[htbp]
	\centering
	\includegraphics[scale=0.50]{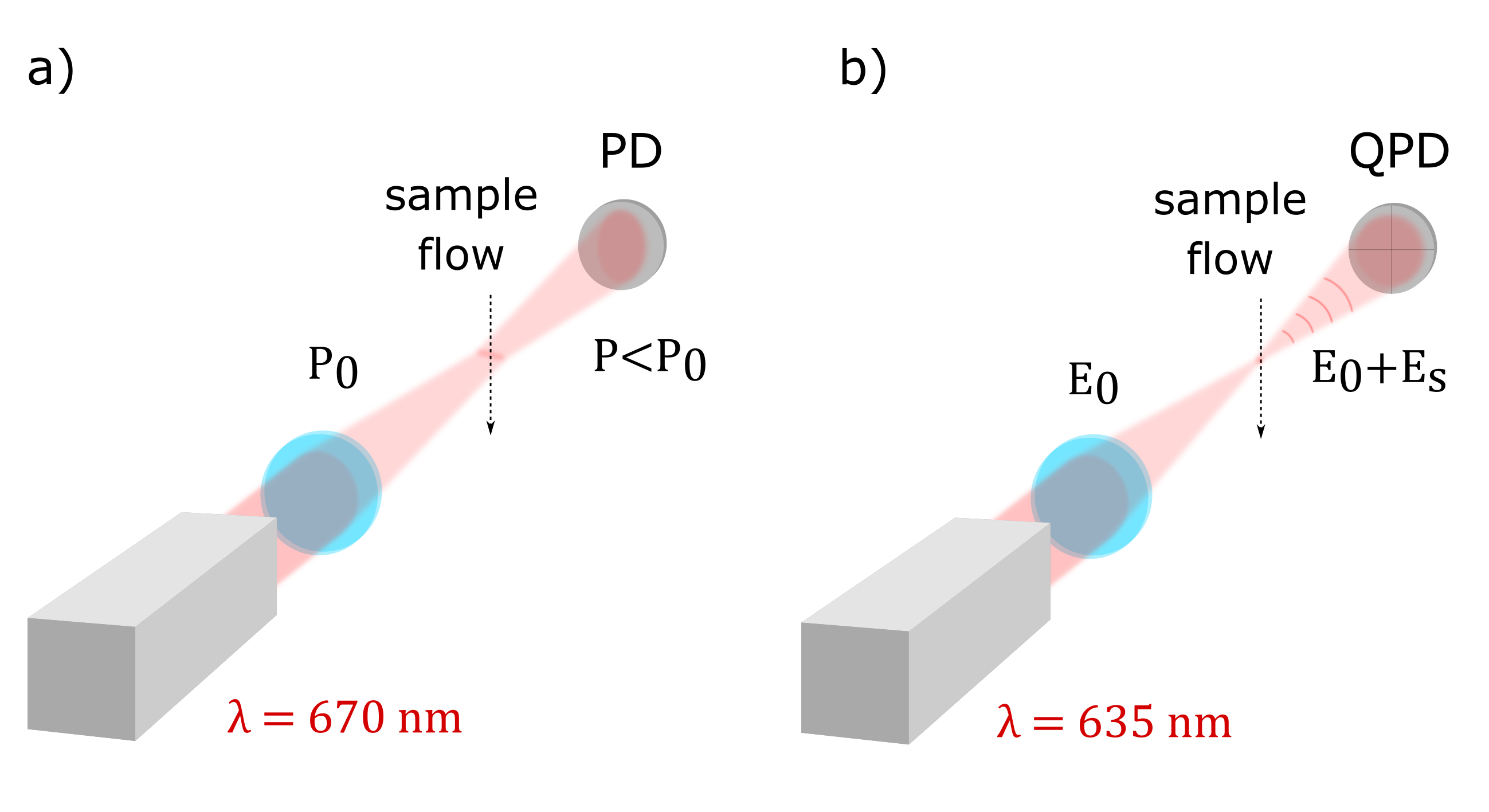} 
	\caption{(a) Schematic of the LO instrument. A \SI{670}{\nano\meter} laser of power $P_0$ impinges perpendicularly on a cell where the sample is flown; when a particle crosses the astigmatic beam focus, a photodiode (PD) measures a decrease in the transmitted power $P$. The particle size is then retrieved from the optical extinction cross-section. (b) Schematic of the SPES method. A \SI{635}{\nano\meter} laser is focused inside a quartz cell; the (interfering) transmitted and scattered fields are collected in the forward direction by a four-quadrant photodiode (QPD). The beam waist is optimized to measure particles with sizes between some hundreds of nanometres and \SI{10}{\micro\meter}, approximately. }
	\label{fig:outline}
\end{figure}

The LO instrument (sketched in Figure \ref{fig:outline}a) we employed for this study is the commercial Abakus particle counter (model LDS23/25bs; Klotz GmbH, Germany). 

The liquid sample is pumped through a small quartz cell illuminated perpendicularly to the liquid flow direction by a laser beam (\textbf{$P_{0}$}) with $\lambda_1=\SI{670}{\nano\meter}$ wavelength \cite{abakus}.
The measuring cell has a cross-section of \SI{250}\times\SI{230}{\micro\meter}. When a microparticle intercepts the beam, a photodiode measures the decrease in the transmitted power (\textbf{P<P$_{0}$}), thus $C_\mathrm{ext}$. The particle size is therefore retrieved by Mie theory \cite{VdH}, based on reference polystyrene spherical particles with refractive index of 1.58(48) at $\lambda_1=\SI{670}{\nano\meter}$. The instrument can measure particles in the range \SI{1}-\SI{10}{\micro\meter}. The bin width has been set to \SI{300}{\nano\meter} for convenience, a prescription that must be taken into account when analyzing LO data.
We compare our LO results with data acquired with a SPES instrument, based on a recently developed optical method \cite{spes_1, spes_2, spes_3}. The layout of this technique is sketched in Figure \ref{fig:outline}b. It relies on the far-field self-reference interference between the zero-angle field scattered by a particle passing through a focused laser beam ($\lambda_2=$ 635 nm), \textbf{E$_S$}, and the (much more intense) field transmitted through the sample, $\mathbf{E}_0$ \cite{spes_1, spes_2}. Its cell has the same geometry (including the cross-section) as that of the LO, which ensures the same fluidic conditions. Low concentration in the samples guarantees the single-particle condition is met, however, internal checks are able to reject possible overlaps or spurious signals. We chose this method because the simultaneous measurement of two independent parameters related to $C_\mathrm{ext}$ and the polarizability of the particle, $\alpha$, makes it possible to obtain quantitative information about both the refractive index and the size of individual particles. The small difference in the operating wavelengths of the two instruments has been considered as regards the refractive index of the samples. The comparatively small difference in the refractive index has been found to be negligible for the correction to the size inversion and can be disregarded for our purposes (see Appendix \ref{appendix_A}). 


\section{Structure of the code}
\label{sec:code}

\begin{figure}[htbp]
	\centering
	\includegraphics[scale=0.28]{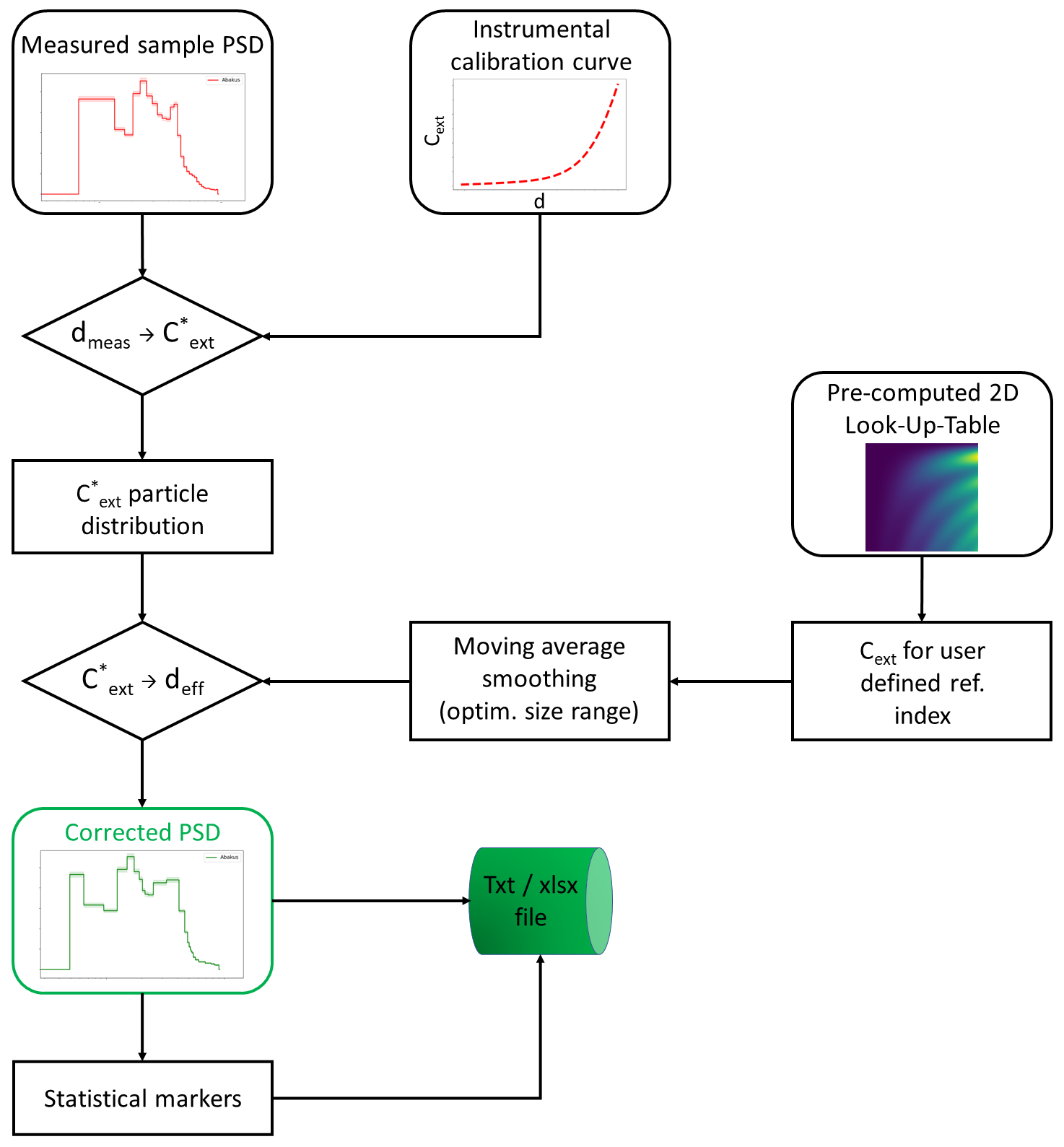} 
	\caption{General workflow of DEDALO.}
	\label{workflow}
\end{figure}

DEDALO is an open-source Python-based GUI software that allows to operate an in-line LO instrument and further analyze the data by compensating for the refractive index according to the sample composition. In addition, DEDALO recovers the numeric concentration of the sample, which is not provided by the instrument. The algorithm computes $C_\mathrm{ext}$ analytically with Mie scattering theory \cite{BH, VdH}; smoothing functions are then used (see below) as customary in traditional instruments. The general workflow is sketched in Figure \ref{workflow} and can be summarized as follows:

\begin{enumerate}[i)]
    \item the instrumental calibration curve is retrieved by measuring mono-disperse suspensions of polystyrene spheres;
    \item the sample PSD is measured;
    \item each bin of the measured PSD is converted into its corresponding $C_\mathrm{ext}$ bin by computing their limits with the calibration curve, thus obtaining a new histogram by counting the number of detected particles within each $C_\mathrm{ext}$ bin;
    \item the lower and upper limits of the $C_\mathrm{ext}$ bins are converted into the corresponding diameters through Mie theory with the new refractive index; in order to save computational resources a pre-computed look-up table is used;
    \item the corrected PSD is written on files (both spreadsheet and text file), together with some statistical markers such as the PSD mode diameter and the distribution quantiles. 
\end{enumerate}

\begin{figure}[htbp]
	\centering
	\includegraphics[scale=0.45]{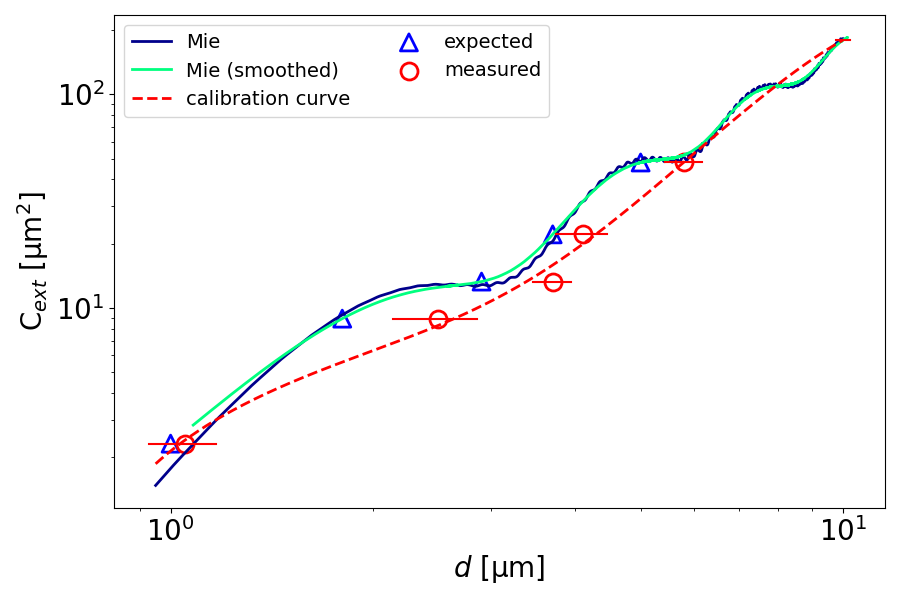} 
	\caption{$C_\mathrm{ext}$ curve from Mie scattering theory (blue solid line) and the smoothed version (green solid line). The LO results for calibrated \SI{1.0}{\micro\meter}, \SI{1.8}{\micro\meter}, \SI{2.9}{\micro\meter}, \SI{3.5}{\micro\meter} and \SI{5.0}{\micro\meter} polystyrene spheres are shown as red circles, compared to the expected $C_\mathrm{ext}$ values (blue triangles). The LO calibration function adopted here to invert $C_\mathrm{ext}$ to $d$, obtained by interpolating the experimental results, is reported as a red dashed line.}
	\label{fig:mie_curves}
\end{figure}

Commercial LO instruments do not typically allow the user to access raw data. A histogram of the particle counts within a given size bin is generated by inverting $C_\mathrm{ext}$ data for each particle through a calibration curve obtained from the Mie function $C_\mathrm{ext}$ \textit{vs} $d$, as discussed for example in \cite{marius}. 
It is convenient to set the instrument in such a way that the width of the bins is uniform throughout the size range. We set the bin width to \SI{300}{\nano\meter}. The calibration curve can be estimated by means of several measurements performed with mono-disperse polystyrene spheres. In Figure \ref{fig:mie_curves}, we report on the $(C_\mathrm{ext},d)$ plane the experimental results of LO measurements (red circles) for \SI{1.0}{\micro\meter}, \SI{1.8}{\micro\meter}, \SI{2.9}{\micro\meter}, \SI{3.5}{\micro\meter} and \SI{5.0}{\micro\meter} calibrated polystyrene spheres and the corresponding expected positions (blue triangles). For each nominal diameter, $C_\mathrm{ext}$ was computed through Mie theory and compared to the particle size reported by the LO instrument. The Mie curve is also reported with a blue solid line, while the green one represents a smoothed Mie curve. The LO calibration curve (red dashed line) has been obtained as a fourth-degree polynomial interpolation of the LO data in this plane:

\begin{equation}
    C(d) = a_0 + a_1 d + a_2 d^2 + a_3 d^3 + a_4 d^4 \quad ,
\end{equation}

where $d$ is the particle diameter in \si{\micro\meter} and $C$ is in \si{\micro\meter\squared}; the degree of the interpolating polynomial, limited by the number of available data points, is the smallest that maximizes compatibility with the real curve. Polynomial coefficients are listed in Table \ref{tab:coefficient}. We stress that the calibration curve does depend on the specific instrument or calibration, so that it must be characterized by measuring calibrated spheres, as discussed above. This step is preliminary to any further operation to recover the $C_\mathrm{ext}$ values corresponding to the diameters limiting the bins of the histogram provided by the instrument. 

\begin{table}[htbp]
\renewcommand{\arraystretch}{1.1}
\centering
\caption{Polynomial coefficients of the LO calibration function.\label{tab:coefficient}}
\smallskip
\begin{tabular}{c|c|c|c|c}
\hline\hline
$a_0$ & \hspace{4mm} $a_1$ \hspace{4mm} & \hspace{4mm} $a_2$ \hspace{4mm} & \hspace{4mm} $a_3$ \hspace{4mm} & \hspace{4mm} $a_4$ \hspace{4mm} \\
\si{\micro\meter\squared} & \si{\micro\meter} & dimensionless & \si{\per\micro\meter} & \si{\per\square\micro\meter} \\
$-4.75\,\cdot10^{-2}$ & +1.02 & $-4.76$ & $+1.21\,\cdot10^{-2}$ & $-6.13$ \\
\hline\hline
\end{tabular}
\end{table}

As a first step, DEDALO converts the lower and upper limits of each size bin into the $C_\mathrm{ext}$ values retrieved through the calibration curve (red dashed). Then, a histogram is obtained by counting the number of detected particles within each $C_\mathrm{ext}$ bin. Finally, the limits of the $C_\mathrm{ext}$ bins are converted again into the corresponding diameters using Mie theory with a refractive index $n$ that best suits the sample under consideration and the histogram values are divided by the corresponding bin width. We introduce a smoothing procedure to damp any local non-monotonic behavior of the Mie functions $C_\mathrm{ext}$ \textit{vs} $d$ with a moving-average within a size range \SI{0.1}{\micro\meter} wide. DEDALO dynamically adjusts the size range of the moving average depending on the desired refractive index. In Figure \ref{fig:mie_curves}, the smoothed function for polystyrene is reported in green.

\begin{figure}[htbp]
	\centering
	\includegraphics[scale=0.42]{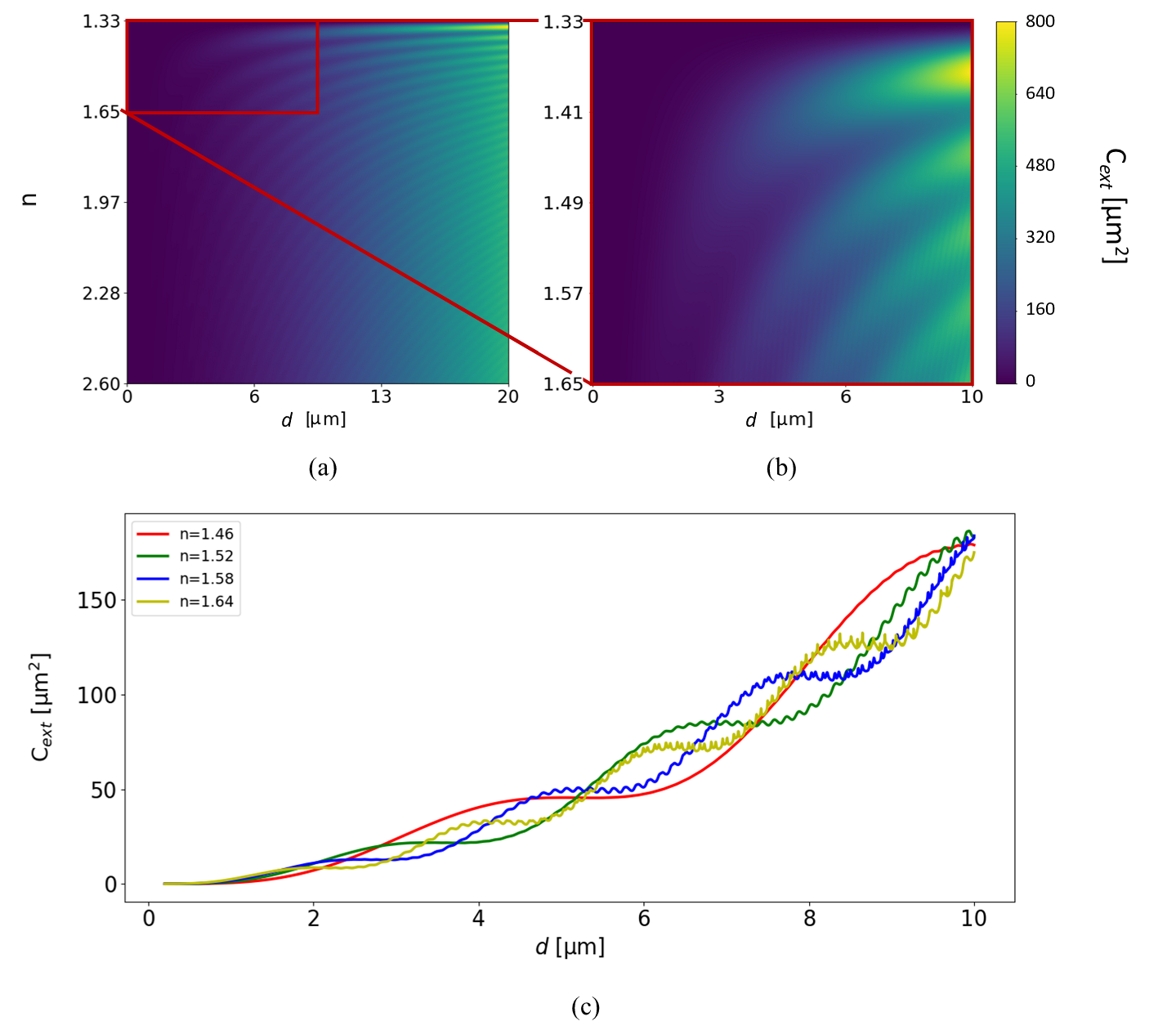} 
	\caption{(a) $C_\mathrm{ext}$ LUT calculated through Mie theory for refractive indices between 1.3311 and 2.6 and particle sizes in the range \SI{0.2}-\SI{20}{\micro\meter}. The inset (b) on the right focuses on the ranges which were most relevant for our work: refractive indices between 1.3311 and 1.6500 and particle sizes ranging from 0.2 to \SI{10}{\micro\meter}. Both images are normalized to their peak value. (c) $C_\mathrm{ext}$ vs $d$ for different refractive indices as retrieved from the LUT. Oscillations in Mie functions leading to a non-monotonic behavior occur at smaller diameters for increasing $n$.}
	\label{fig:2D_cext}
\end{figure}

The core of the DEDALO algorithm is a pre-computed $(C_\mathrm{ext},d)$ look-up table (LUT, Figure \ref{fig:2D_cext}a-b) obtained by varying particle diameter and refractive index and calculating the $C_\mathrm{ext}$ values through Mie theory. The use of the LUT instead of recomputing Mie functions saves a considerable amount of computational time. 

\begin{figure}[htbp]
	\centering
	\includegraphics[scale=0.45]{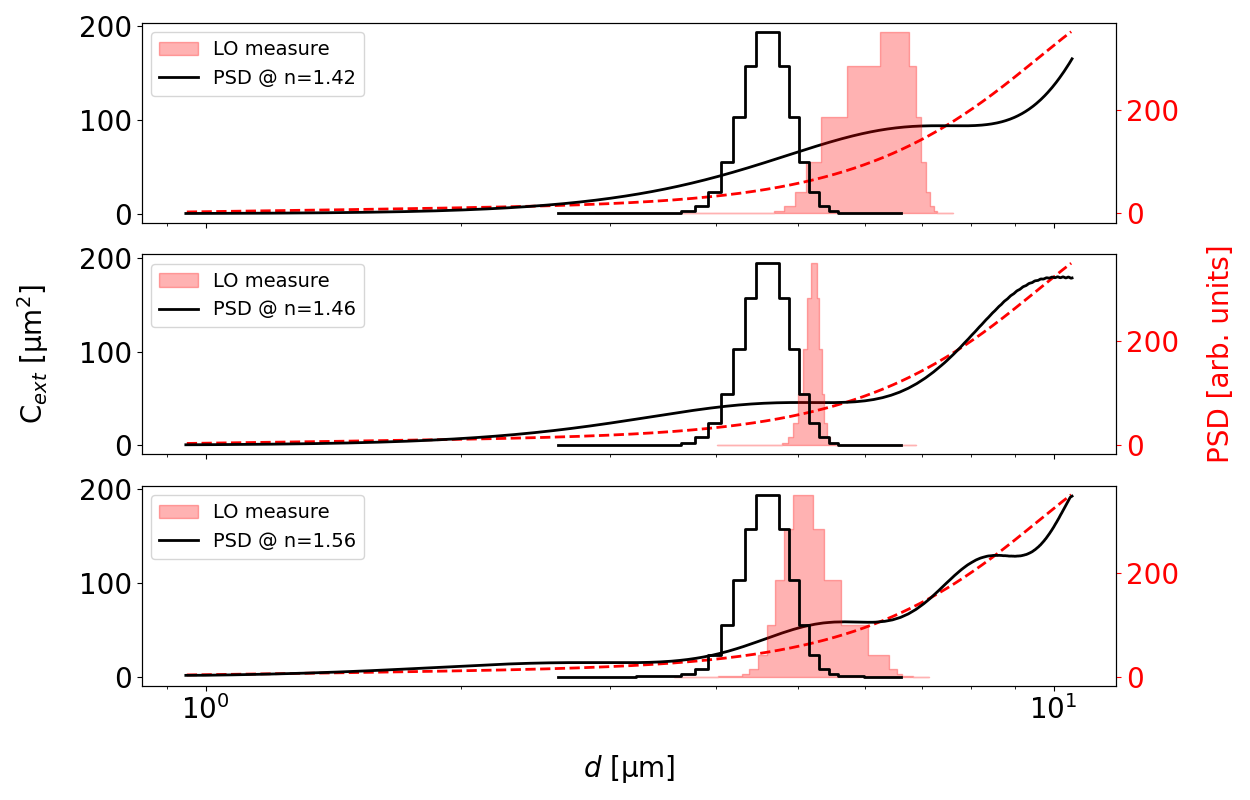}
	\caption{Example of the PSD conversion process. Black solid lines represent the $C_\mathrm{ext}$ curves for $n=1.42$, $n=1.46$ and $n=1.56$ respectively, according to Mie theory. The red dashed curve is the LO calibration function. The red histograms are the result of unprocessed LO measurements, whereas the black and white histograms are the effective initial Gaussian PSDs. DEDALO processes the red histograms to give the correct ones. }
	\label{fig:algorithm_correction}
\end{figure}

For any given $n$, a size is associated through the LUT to each $C_\mathrm{ext}$ bin, hence the corresponding histogram value. The LUT has been calculated over a size range of \SI{0.2}-\SI{20}{\micro\meter} and a refractive index range of $1.3311-2.6$. The minimum value of $n$ is determined by that of pure water at the operative wavelength of \SI{670}{\nano\meter} ($n_\mathrm{H_{2}O}=1.3310$). 
In Figure \ref{fig:algorithm_correction}, the algorithm working procedure is shown for hypothetical samples with $n=1.42$, $n=1.46$ and $n=1.56$, all of them characterized by a Gaussian PSD with average \SI{4.5}{\micro\meter} and variance \SI{0.3}{\micro\meter} (black and white histograms). Red histograms are the hypothetical results of measurements providing the PSD obtained through LO. DEDALO allows to recover the effective Gaussian histograms from them. The red dashed line represents the calibration curve, adopted here to convert $d$ into $C_\mathrm{ext}$. The black solid curves are obtained from the LUT for the considered refractive index, to convert $C_\mathrm{ext}$ into $d$. As discussed in more detail below, the remarkable difference between the distributions proves the need to consider the true refractive index to avoid systematic errors in the PSDs. Following the manufacturer's indications, the instrumental calibration curve should be characterized periodically and each time the instrument is subject to relevant changes that could affect the results. 

DEDALO allows the user to operate the LO instrument in-line, requiring only the value of the flow rate as an input. This is just to calculate particle concentrations based on particle counts. During the measurement, DEDALO shows the instantaneous (\SI{1}{\second} integration time) and the time-integrated PSD, as well as the instantaneous numeric concentration of the sample. A continuous monitoring of the working parameters is provided. The instantaneous and time-integrated PSDs are shown in the interface and are written into a file, together with some statistical markers such as the PSD mode diameter and the distribution quantiles. For convenience, the data is copied into a spreadsheet file as well as a plain text file, to comply with most data visualization tools. 


\section{Results}
\label{sec:data}

The DEDALO algorithm proved to be effective in measuring samples with refractive indices ranging from 1.46 to 1.64, using emulsions as a case study. We also tested the system by mixing two emulsions with known and controlled volume fractions, once analyzed separately. While absorption was negligible for all the analyzed samples, we note that it is not a strict requirement of the method nor the protocol. In fact, in case of absorbing particles, DEDALO can still account for absorption by changing the inversion curve. The choice to operate with samples such as those shown in Table \ref{tab:emulsions} stems mainly from the intention to develop and validate DEDALO under the best possible operating conditions for the LO device, i.e. when using dielectric materials and the spherical approximation applies. Conversely, absorbing nano- and micro-particles are typically characterized by non-spherical shapes and a rather irregular morphology.

The samples were flown into the LO and SPES instruments with a constant flow rate of 2 mL/min; the two instruments were connected in series, thus, the samples can be safely considered statistically the same. The flow has been chosen to minimize any possible deformation of the oil droplets due to shear. While it was not possible to set flow rates lower than 1 mL/min due to the limits of the LO measuring capabilities, we have verified that the chosen flow rate does not significantly affect the spherical geometry thanks to the measurement of two independent parameters obtained simultaneously by the SPES instrument. \cite{universality_lscatter}. Moreover, SPES provides $C_\mathrm{ext}$ histogram as raw data, giving a unique check to the output of DEDALO from each LO measurement. 

\begin{figure}[htbp]
	\captionsetup[subfigure]{labelformat=empty}
	\begin{subfigure}{.5\textwidth}
	\centering
	\hspace{.6cm}\includegraphics[scale=0.4]{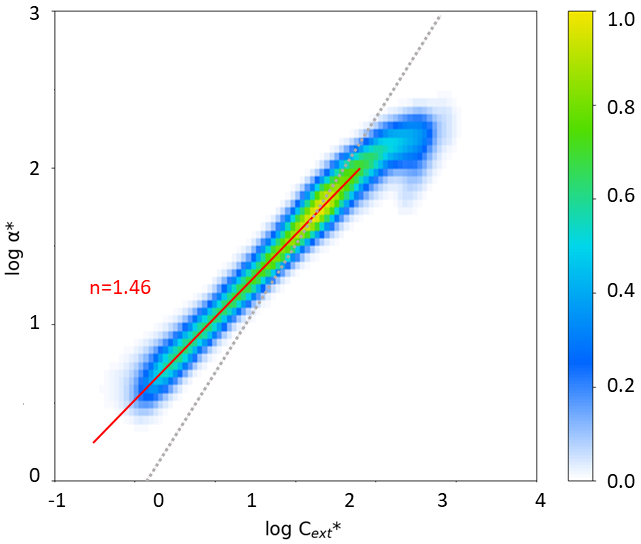} 
	\caption{(a) $n=1.46$ SPES measurement}
	\end{subfigure}
\begin{subfigure}{.5\textwidth}
	\centering	
	\hspace{.6cm}\includegraphics[scale=0.4]{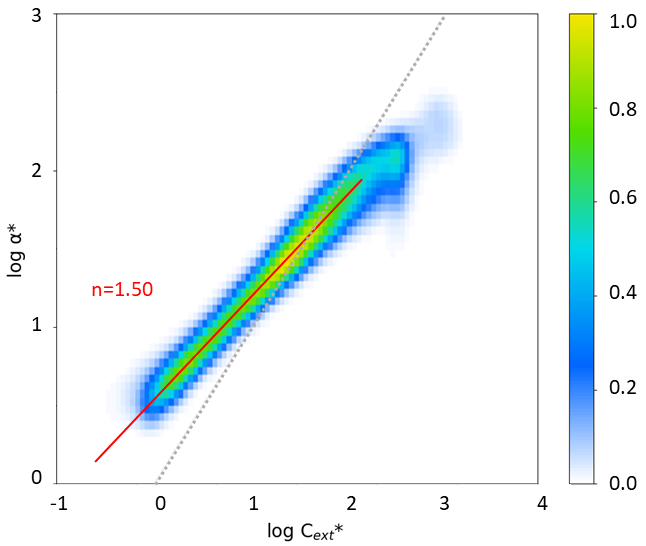} 
	\caption{(b) $n=1.50$ SPES measurement}
	\end{subfigure}
\\[2ex]
\begin{subfigure}{.5\textwidth}
	\centering	
	\includegraphics[scale=0.315]{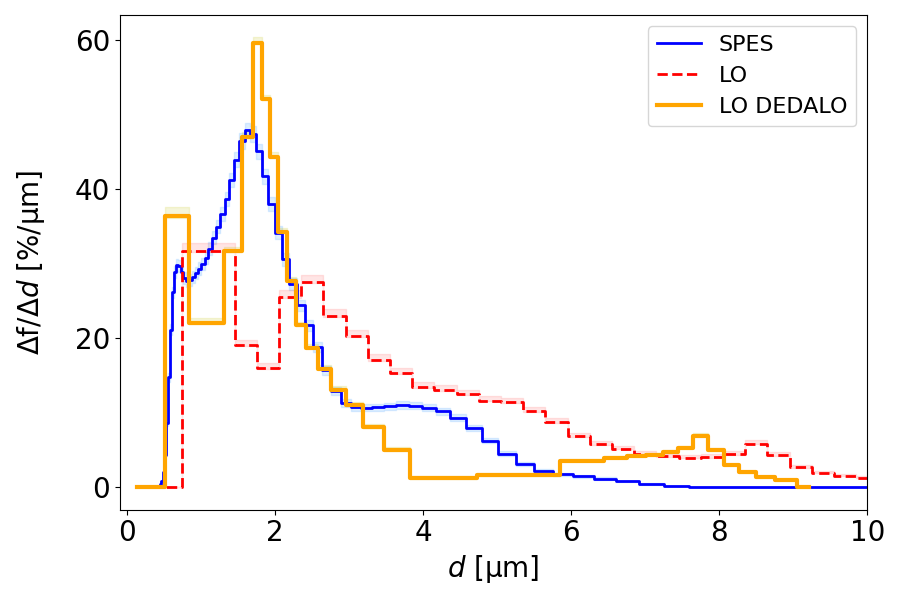} 
	\caption{(c) $n=1.46$ before DEDALO}
	\end{subfigure}
    \begin{subfigure}{.5\textwidth}
	\centering	
	\includegraphics[scale=0.315]{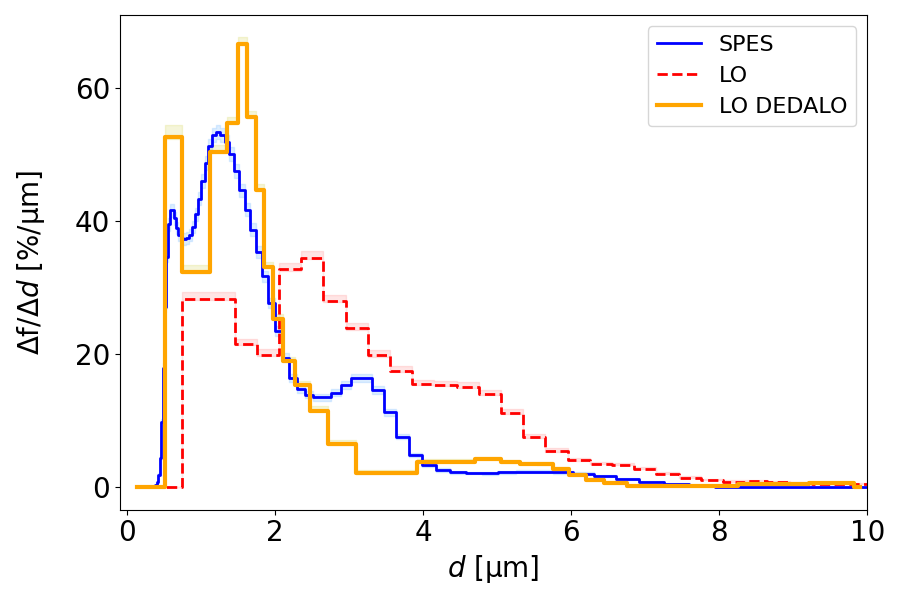} 
	\caption{(d) $n=1.50$ before DEDALO}
	\end{subfigure}
	\caption{Results obtained with $n=1.46$ and $n=1.50$ refractive index liquid emulsions. (a-b) SPES data allow to retrieve the refractive index by fitting Mie curves to the 2D histograms. The grey dotted line indicates the diagonal of the plane as a guide to the eye. (c-d) the PSD as measured with SPES (blue histogram, logarithmic sampling), LO (red dashed histogram, linear sampling), LO data processed with DEDALO (solid orange histogram). Color bands indicate uncertainties, calculated one standard deviation per bin.}
	\label{fig:results1}
\end{figure}

\begin{figure}[htbp]
	\captionsetup[subfigure]{labelformat=empty}
\begin{subfigure}{.5\textwidth}
	\centering
    \includegraphics[scale=0.315]{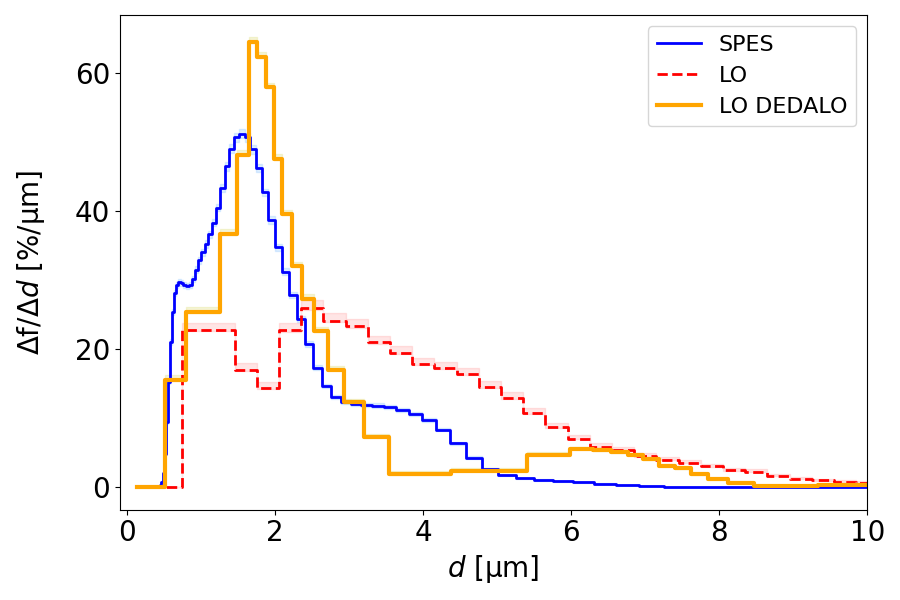} 
	\caption{(a) $n=1.47$}
	\end{subfigure} 
\begin{subfigure}{.5\textwidth}
	\centering	
	\includegraphics[scale=0.315]{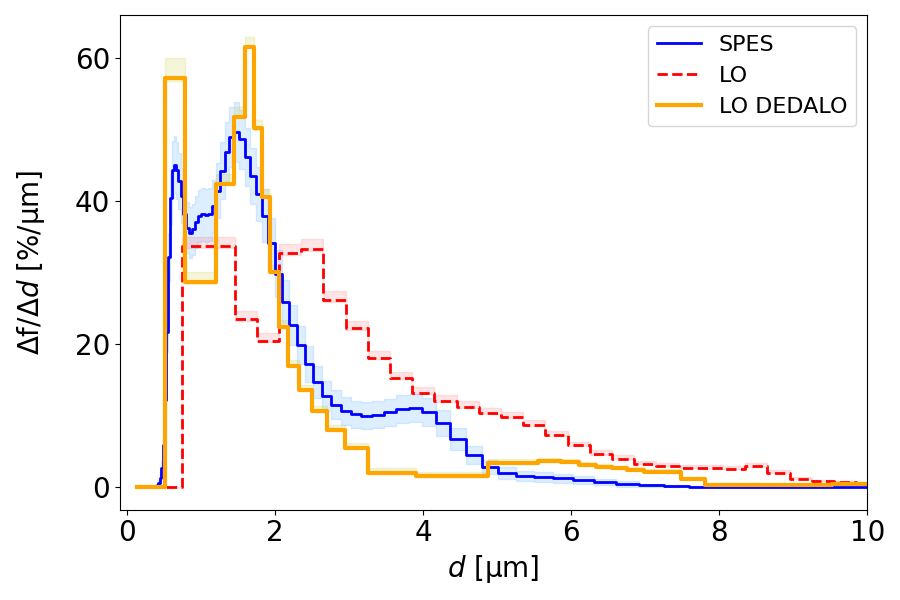} 
	\caption{(b) $n=1.48$}
	\end{subfigure} 
\begin{subfigure}{.5\textwidth}
	\centering
	\includegraphics[scale=0.315]{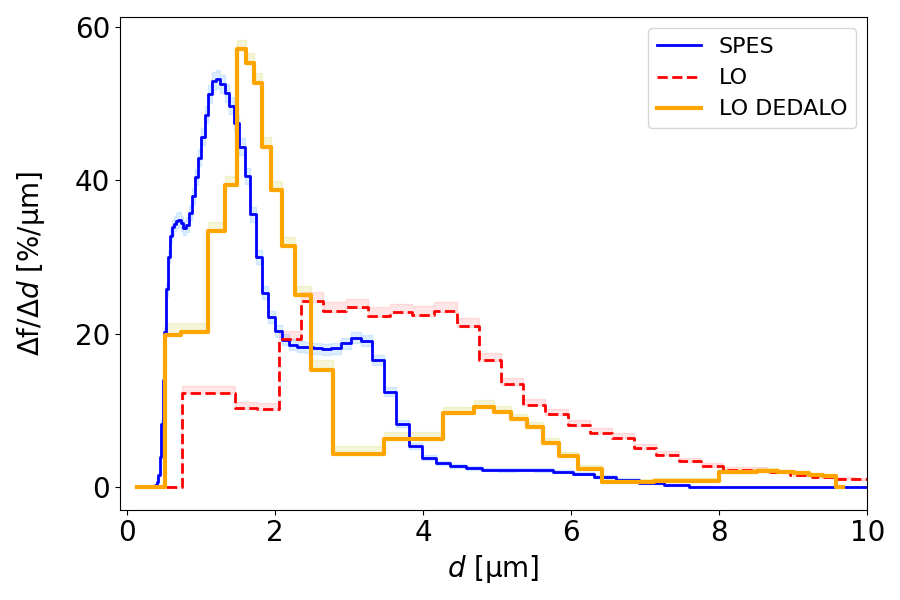} 
	\caption{(c) $n=1.51$}
	\end{subfigure} 
\begin{subfigure}{.5\textwidth}
	\centering	
	\includegraphics[scale=0.315]{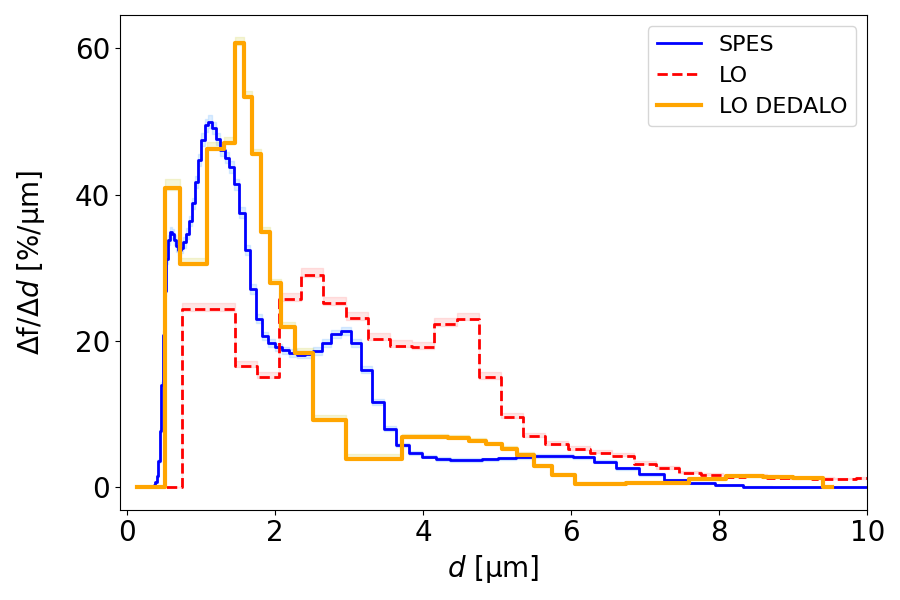} 
	\caption{(d) $n=1.52$}
	\end{subfigure} 
	\caption{PSDs measured by LO device and obtained through DEDALO from these measurements (red dashed histograms and solid orange histograms respectively, linear sampling) compared to the corresponding SPES data (blue histograms, logarithmic sampling) for oil emulsions with refractive index (a) 1.47, (b) 1.48, (c) 1.51 and (d) 1.52. Color bands correspond to uncertainties calculated as the square root of the number of particles measured per bin and expressed as 1 standard deviation.}
	\label{fig:all1}
\end{figure}

\begin{figure}[htbp]
	\captionsetup[subfigure]{labelformat=empty}
\begin{subfigure}{.5\textwidth}
	\centering
	\includegraphics[scale=0.315]{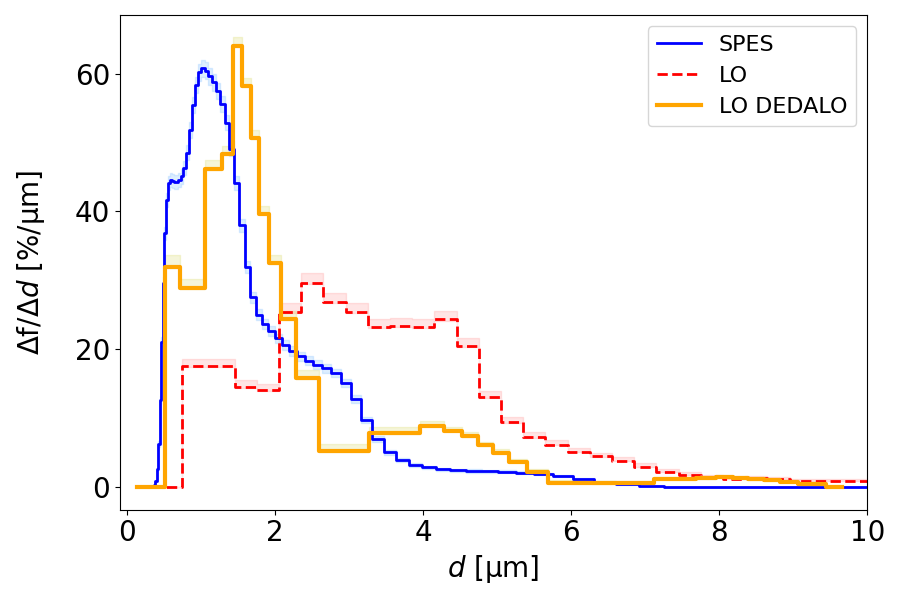} 
	\caption{(a) $n=1.53$}
	\end{subfigure} 
\begin{subfigure}{.5\textwidth}
	\centering	
	\includegraphics[scale=0.315]{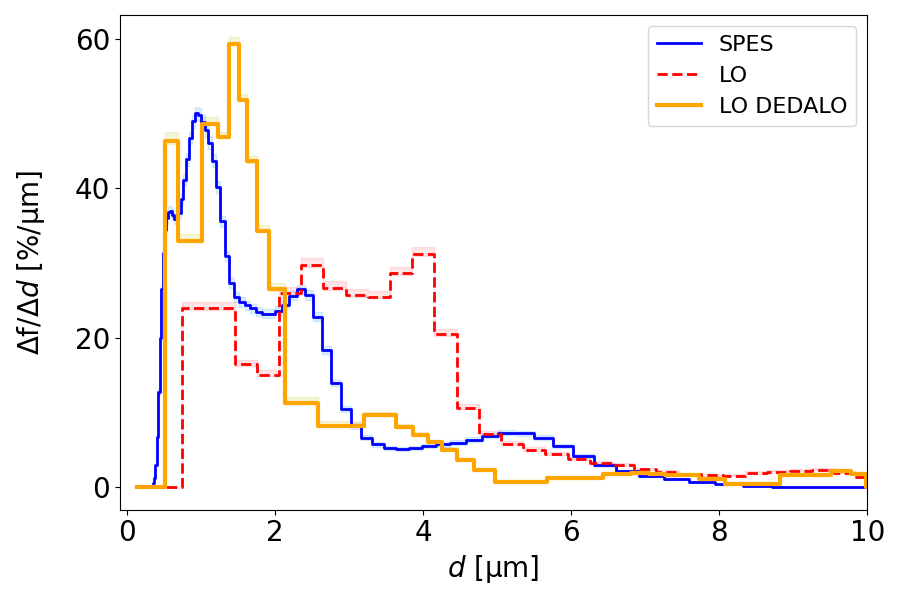} 
	\caption{(b) $n=1.56$}
	\end{subfigure} 
\begin{subfigure}{.5\textwidth}
	\centering
	\includegraphics[scale=0.315]{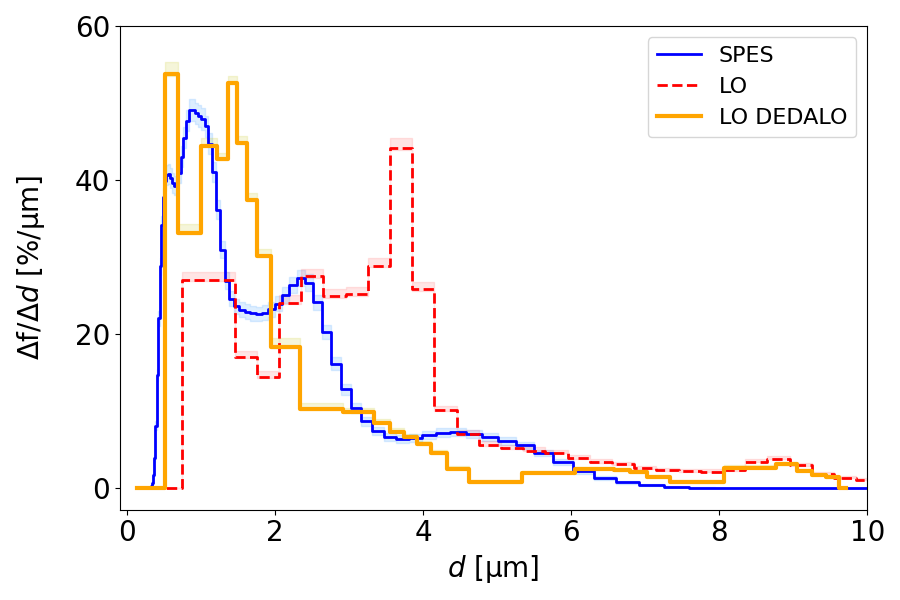} 
	\caption{(c) $n=1.58$}
	\end{subfigure}  
\begin{subfigure}{.5\textwidth}
	\centering	
	\includegraphics[scale=0.315]{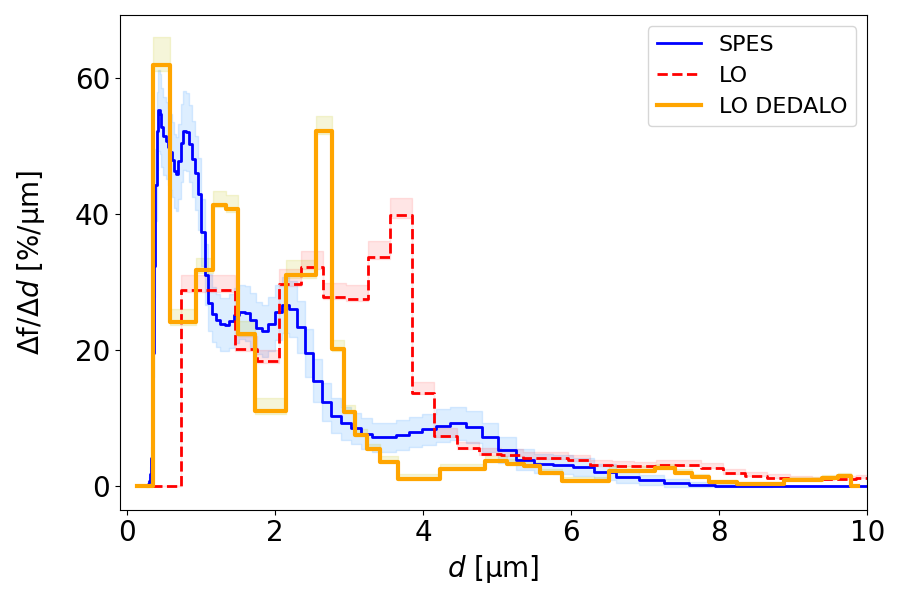} 
	\caption{(d) $n=1.64$}
	\end{subfigure}
	\caption{PSDs measured by LO device and obtained through DEDALO from these measurements (red dashed histograms and solid orange histograms respectively, linear sampling) compared to the corresponding SPES data (blue histograms, logarithmic sampling) for oil emulsions with refractive index (a) 1.53, (b) 1.56, (c) 1.58 and (d) 1.64. Considering also the results reported in the previous figure, it is evident that the larger the sample refractive index, the greater the disagreement between SPES and DEDALO corrected distributions. With regard to uncertainties, the same applies as above.}
	\label{fig:all2}
\end{figure}

\begin{figure}[htbp]
	\captionsetup[subfigure]{labelformat=empty}
\begin{subfigure}{.5\textwidth}
	\centering
	\hspace{.6cm}\includegraphics[scale=0.38]{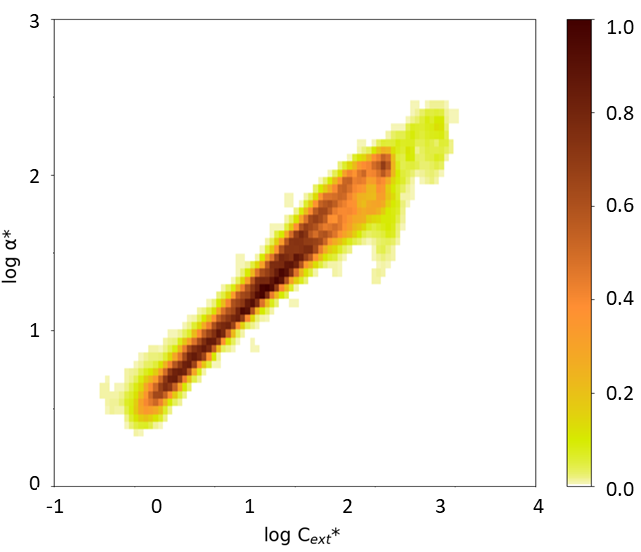} 
    \caption{(a) $n_1=1.50$ and $n_2=1.52$ SPES measurement}
	\end{subfigure} \hspace{0.01cm}
\begin{subfigure}{.5\textwidth}
	\centering	
	\hspace{.6cm}\includegraphics[scale=0.38]{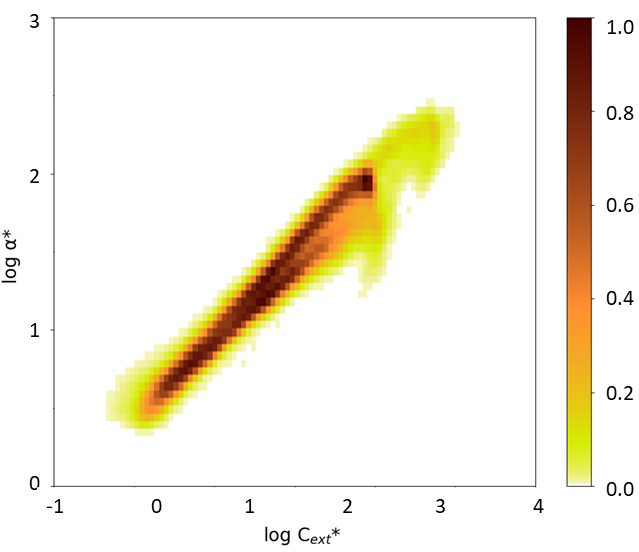} 
    \caption{(b) $n_1$=1.50 and $n_2$=1.56 SPES measurement}
	\end{subfigure}
\\[2ex]
\begin{subfigure}{.5\textwidth}
	\centering
	\includegraphics[scale=0.295]{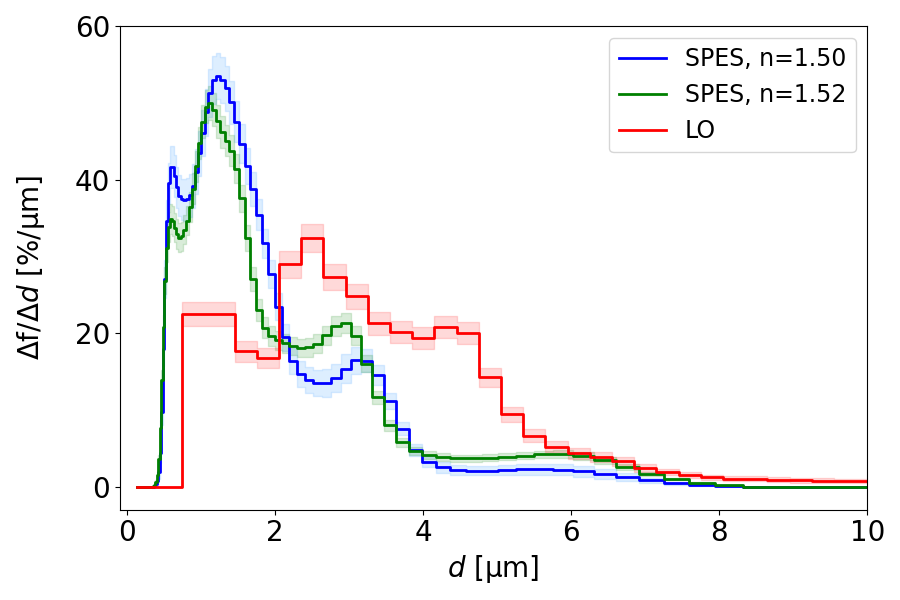} 
    \caption{(c) $n_1$=1.50 and $n_2$=1.52 before DEDALO}
	\end{subfigure} \hspace{0.01cm}
\begin{subfigure}{.5\textwidth}
	\centering	
	\includegraphics[scale=0.295]{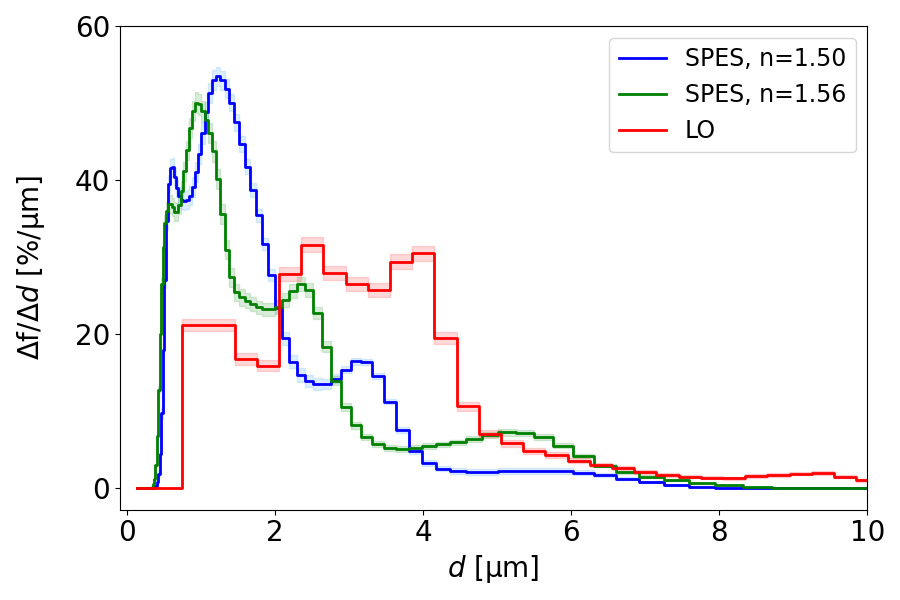} 
    \caption{(d) $n_1$=1.50 and $n_2$=1.56 before DEDALO}
	\end{subfigure}
\\[2ex]
 \begin{subfigure}{.5\textwidth}
	\centering
	\includegraphics[scale=0.295]{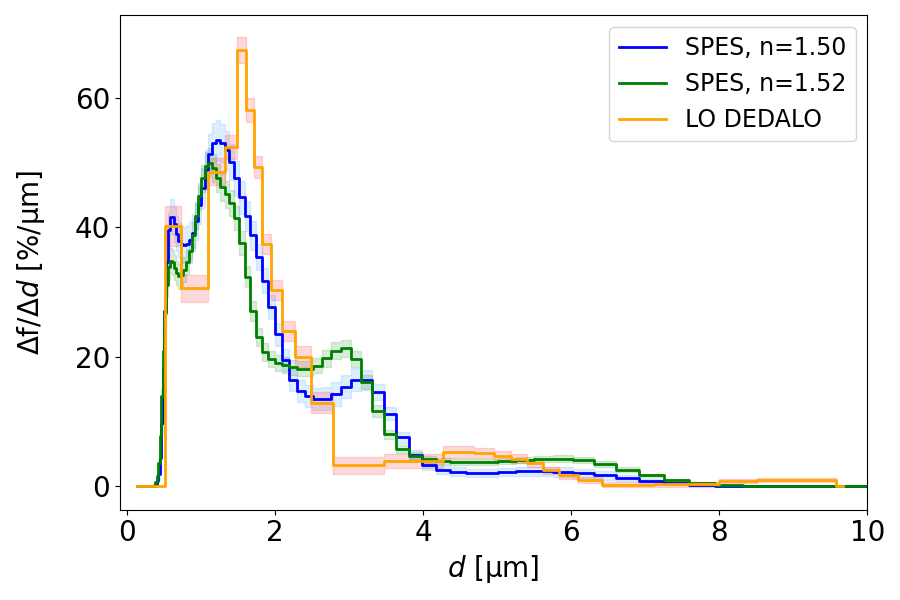}
    \caption{(e) $n_1$=1.50 and $n_2$=1.52 after DEDALO}
	\end{subfigure} \hspace{0.01cm}
\begin{subfigure}{.5\textwidth}
	\centering	
	\includegraphics[scale=0.295]{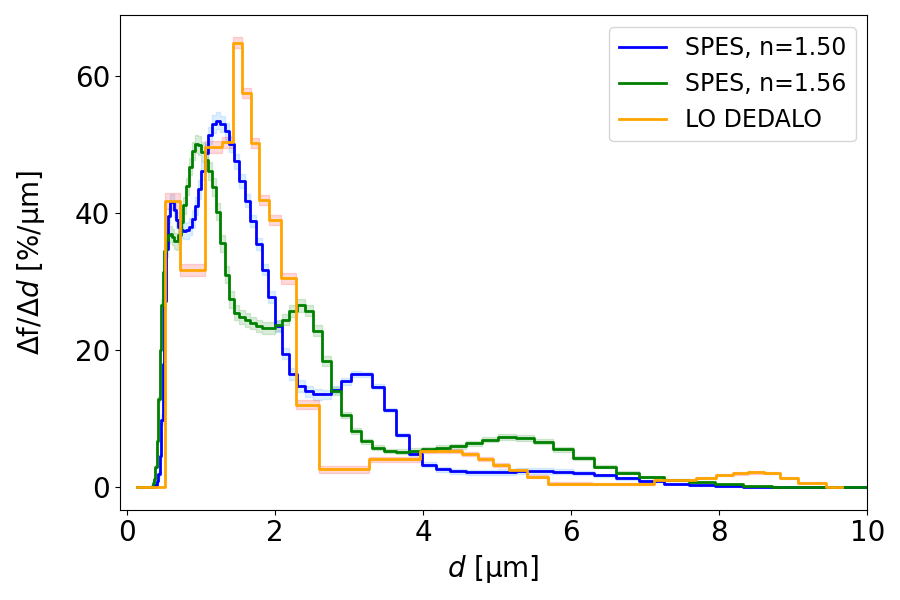} 
	\caption{(f) $n_1$=1.50 and $n_2$=1.56 after DEDALO}
	\end{subfigure}
	\caption{Examples of the results obtained with mixed refractive indices emulsions: $n=1.50$ \& $n=1.52$ (on the left side) and $n=1.50$ \& $n=1.56$ (on the right side). (a-b) The two constituent oils, both characterized by a different refractive index value from the polystyrene calibration standard, are correctly distinguished by SPES technique. (c-f) Here, the PSD measured by the LO device was inverted considering an effective refractive index as the average of the refractive indices of each component. Both before (red histograms) and after (orange histograms) DEDALO execution, the comparison is carried out between the LO data and the PSDs of the two constituent oil emulsions each separately measured by SPES (blue and green histograms, referred to the emulsions with the lower and the higher refractive index values respectively). With regard to sampling, the same applies as above.}
	\label{fig:results2}
\end{figure}

Figures \ref{fig:results1}-\ref{fig:results2} report some of the results we obtained with single and mixed oil emulsion, respectively.
In Figure \ref{fig:results1}a-b SPES experimental results are shown as 2D histograms reporting the relative number of particles measured within each 2D bin, as a function of the dimensionless extinction cross-section ($C_\mathrm{ext}^*=C_\mathrm{ext}k^2$/4$\pi$) and polarizability ($\alpha^*=\alpha k^3$), $k$ being the wavenumber. Besides using SPES as a double-check for the particle sizing, we could also determine the refractive index of the samples independently, by fitting Mie curves to the experimental data. The agreement with the expected values of $n$ is good for all the samples: the standard deviation ranges from $\sigma_n=$ 2$\cdot$10$^{-2}$ in the case of single oil emulsion up to $\sigma_n=$ 4$\cdot$10$^{-2}$ referring to emulsions obtained by mixing two oils with different $n$.
This is a further indication that the particles behave as uniform spheres. 

Figures \ref{fig:results1}c-d report the PSDs measured by SPES (blue histograms) and LO, both before and after processing data through DEDALO (red dashed histogram and orange histogram respectively). The PSDs are obtained by dividing the counts by the corresponding bin width \cite{hinds}. Finally, PSDs are normalized to the total number of detected particles. Thus, the area of each rectangle is equal to the fraction ($\%$) of particles in the corresponding bin. After our correction, the PSDs obtained from LO data appear in much more agreement with those obtained from SPES. The uncorrected PSDs are shifted toward larger diameters: this is due to the specific behavior of the calibration curve adopted by the LO instrument. It is worth noting that, by considering samples with increasing refractive index (see Figure \ref{fig:all1} and Figure \ref{fig:all2}), we observe an increasing disagreement. This is due to the non-monotonic behavior of the Mie functions, occurring at smaller diameters for increasing $n$, as evident from Figure \ref{fig:2D_cext}b-c. 

A remarkable artifact appears for $d=$ 1 \SI{}{\micro\meter} where the recovered PSDs (orange histograms) appear extremely underestimated compared with SPES data (blue histograms). This artifact is due to the specific behavior of the $C_\mathrm{ext}$ in that size region and the corresponding calibration curve. As it is evident in Figure \ref{fig:mie_curves}, the extinction curve is steeper than the calibration, thus introducing the artifact represented by a minimum around $d=$ 2 \SI{}{\micro\meter} in all the original LO curves (red dashed). The current implementation of the code does not compensate for these artifacts. 

In Figure \ref{fig:results2} we show the results obtained by mixing two emulsions with different refractive indices, a typical real-case scenario in most real-world applications. Since LO devices access one parameter ($C_\mathrm{ext}$) no way is there to distinguish different species in samples composed of several materials. On the contrary, in Figure \ref{fig:results2}a-b SPES clearly shows the two populations of oil spheres composing the emulsion, thus evidencing the unavoidable artifacts when inverting with one index. Nevertheless, by knowing the refractive indices we can attempt to compensate for the difference between the given oils and polystyrene calibration. 
In the case of a mixture of suspensions, we propose to use an effective refractive index.
In Figure \ref{fig:results2}c-d we compare the PSDs of each mixed emulsion as measured by LO (red histogram) and the PSDs of the two constituent oil emulsions separately measured by SPES (blue and green histograms, respectively the emulsions with the lower and the higher refractive index). The same PSDs from the SPES instrument are also reported in Figure \ref{fig:results2}e-f with the PSD obtained through DEDALO (orange histograms) by inverting LO data considering an effective refractive index as the average of the refractive indices of each component.

These results show that, despite the use of DEDALO to account for the effective refractive index of the mixed emulsion, the PSD is less accurate than that obtained from the single oil emulsions obtained by SPES. Still, DEDALO will always be more reliable in terms of PSDs than the unprocessed LO.


\section{Discussion}
\label{sec:discussion}

To quantify the reliability of DEDALO, a statistical approach has been considered. The Pearson correlation coefficient between SPES and LO PSDs before and after the algorithm application was computed as

\begin{equation}
    r_{12} = \dfrac{\sum\limits_{i=1}^{N} \left( h_1(i) - \overline{h_1} \right) \left( h_2(i) - \overline{h_2} \right)}{\sqrt{\sum\limits_{i=1}^N \left( h_1(i) - \overline{h_1} \right)^2} \sqrt{\sum\limits_{i=1}^N \left( h_2(i) - \overline{h_2} \right)^2}} \quad ,
    \label{eq:pearson_eq}
\end{equation}

where $h_1$ and $h_2$ define the PSD referring to SPES and LO respectively and the sum is carried out over the bins of each histogram. $\overline{h_1}$ and $\overline{h_2}$ are the average values of the corresponding size distributions, according to the definition

\begin{equation}
    \overline{h_k} = \dfrac{1}{N} \sum\limits_{k=1}^N h_k(i) ,\hspace*{1cm} k=\{1, 2\} \quad .
    \label{eq:hist_normalization}
\end{equation}

This correlation coefficient defines the ratio between the covariance of the two PSDs and the product of their standard deviations; thus, it is essentially a normalized measurement of their covariance, such that the result always has a value between $-1$ (fully anti-correlated variables) and 1 (fully correlated variables). 

\begin{figure}[htbp]
	\centering
	\includegraphics[scale=0.45]{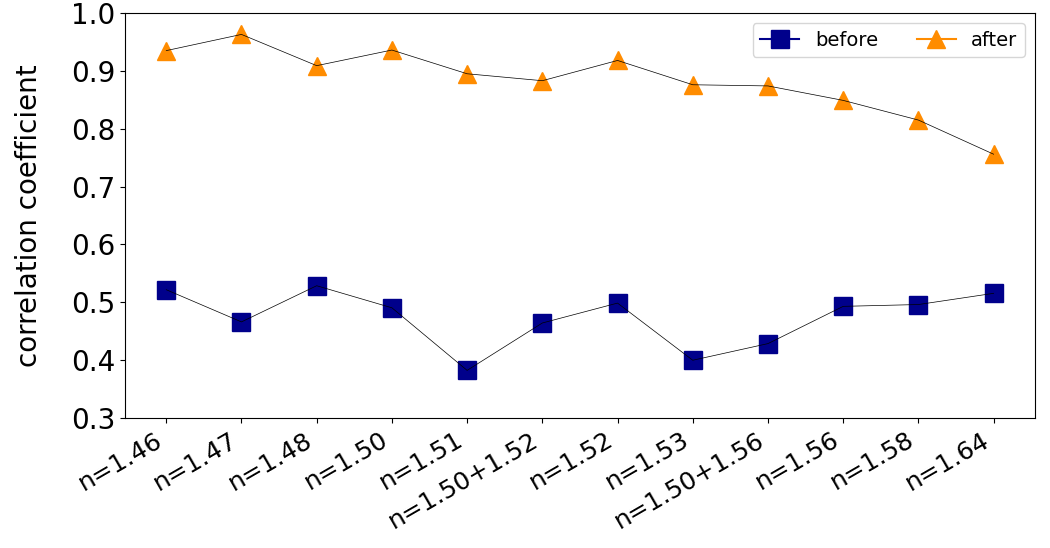} 
	\caption{Pearson correlation coefficient evaluated for all the emulsions considered in this work before and after DEDALO algorithm application (blue squares and orange triangles respectively). Overall, the correlation increases by at least a factor of 2.2. Abscissas are the sample refractive indices, highlighting the progressive worsening of the data reduction as the refractive index increases.}
	\label{fig:pearson}
\end{figure}

Figure [\ref{fig:pearson}] shows the correlation coefficient between SPES and LO instruments PSDs before (blue squares) and after (orange triangles) DEDALO execution for all the tested emulsions.

\begin{figure}[htbp]
	\centering
	\includegraphics[scale=0.45]{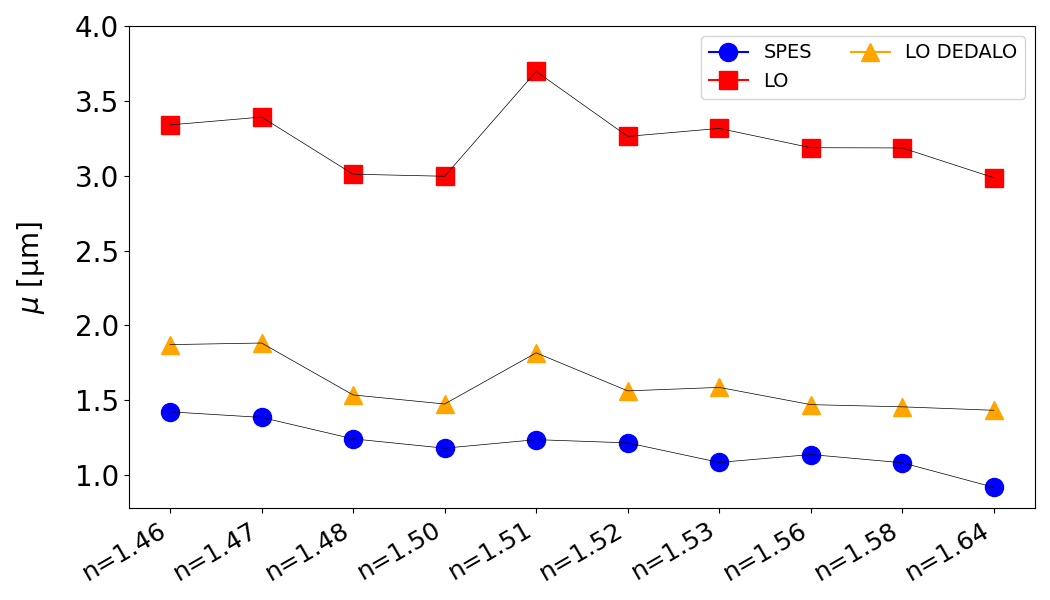} 
	\caption{Geometric median of the PSDs measured by SPES instrument (blue circles) and by the LO device both before (red squares) and after (orange triangles) DEADALO algorithm.}
	\label{fig:geometric_median}
\end{figure}

It is possible to appreciate a considerable improvement (correlation-wise) between the data directly recorded by the LO instrument and those processed by DEDALO. Our results show a good correlation for almost all the indices $\gtrsim$ 0.8 while the original LO data are limited below 0.5. The algorithm described in this work considerably improves particle sizing accuracy and measurement quality, as well as the agreement with the data from other light scattering techniques such as the one considered here. As expected, the larger the difference between the target refractive index and that of the polystyrene calibration standard, the larger the correction. Furthermore, in the case of low values of the refractive index, Mie scattering theory gives C$_\mathrm{ext}$ curves that are considerably smoother and more monotonic. Therefore, the inversion is inherently less prone to artifacts.
The effectiveness of DEDALO is further confirmed by the strongly improved comparability between the PSDs geometric median ($\mu$) after the algorithm processing. By describing PSDs as log-normal distributions, the geometric median can be expressed as $\mu = 10^{\mu*}$, where $\mu*$ defines the average particle diameter weighted on the histogram bin values. These results are reported in Figure \ref{fig:geometric_median}. 

The LO instrument gives a median diameter clearly overestimated compared to the SPES measurement (by more than a factor of 2). The same data processed using DEDALO are much closer to the expected values, within 25$\%$ for all considered samples.

\begin{figure}[htbp]
	\captionsetup[subfigure]{labelformat=empty}
\begin{subfigure}{.5\textwidth}
	\centering
	\includegraphics[scale=0.315]{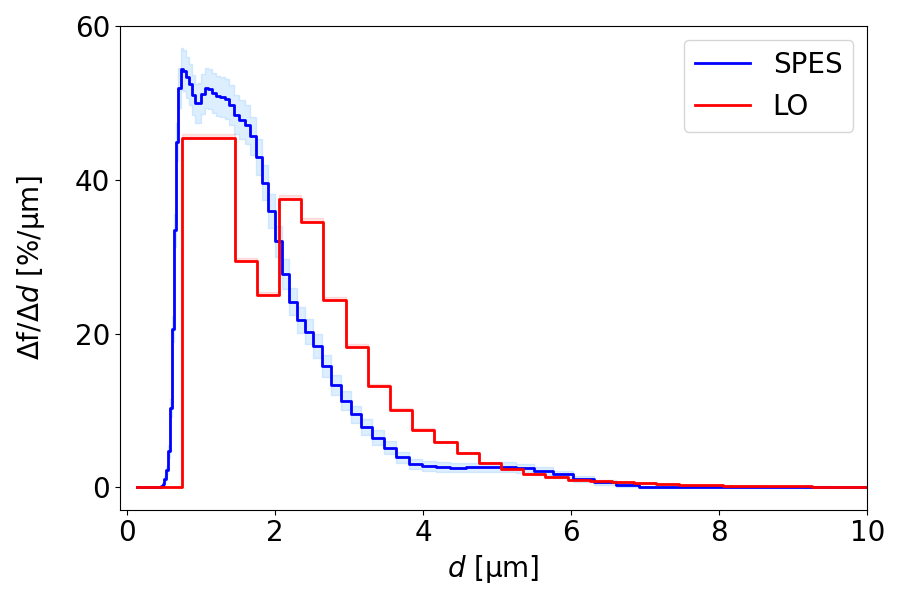}
    \caption{(a) PSDs before DEDALO}
	\end{subfigure} \hspace{0.01cm}
\begin{subfigure}{.5\textwidth}
	\centering	
	\includegraphics[scale=0.315]{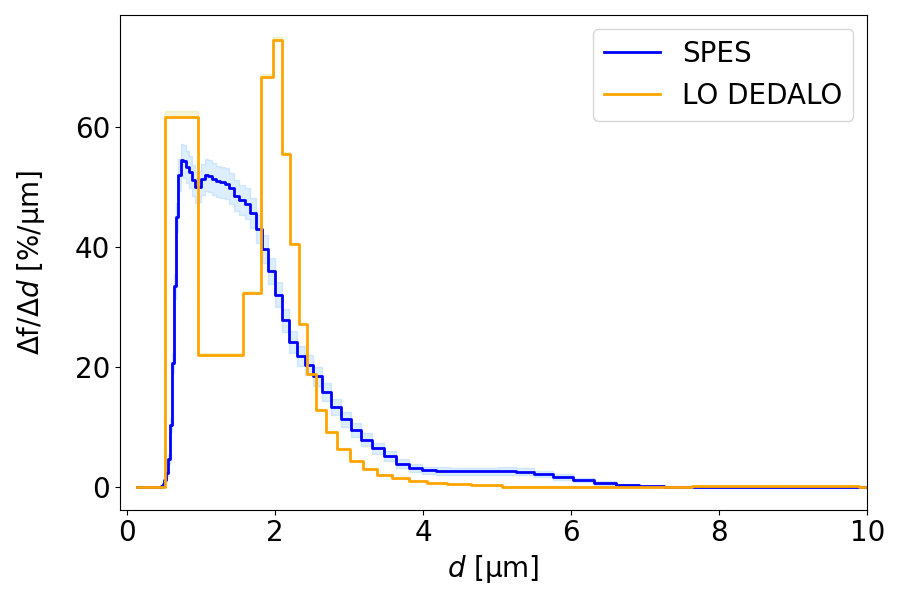} 
    \caption{(b) PSDs after DEDALO}
	\end{subfigure}
\\[2ex]
\begin{subfigure}{1.\textwidth}
	\centering	
	\hspace{0.2cm}\includegraphics[scale=0.325]{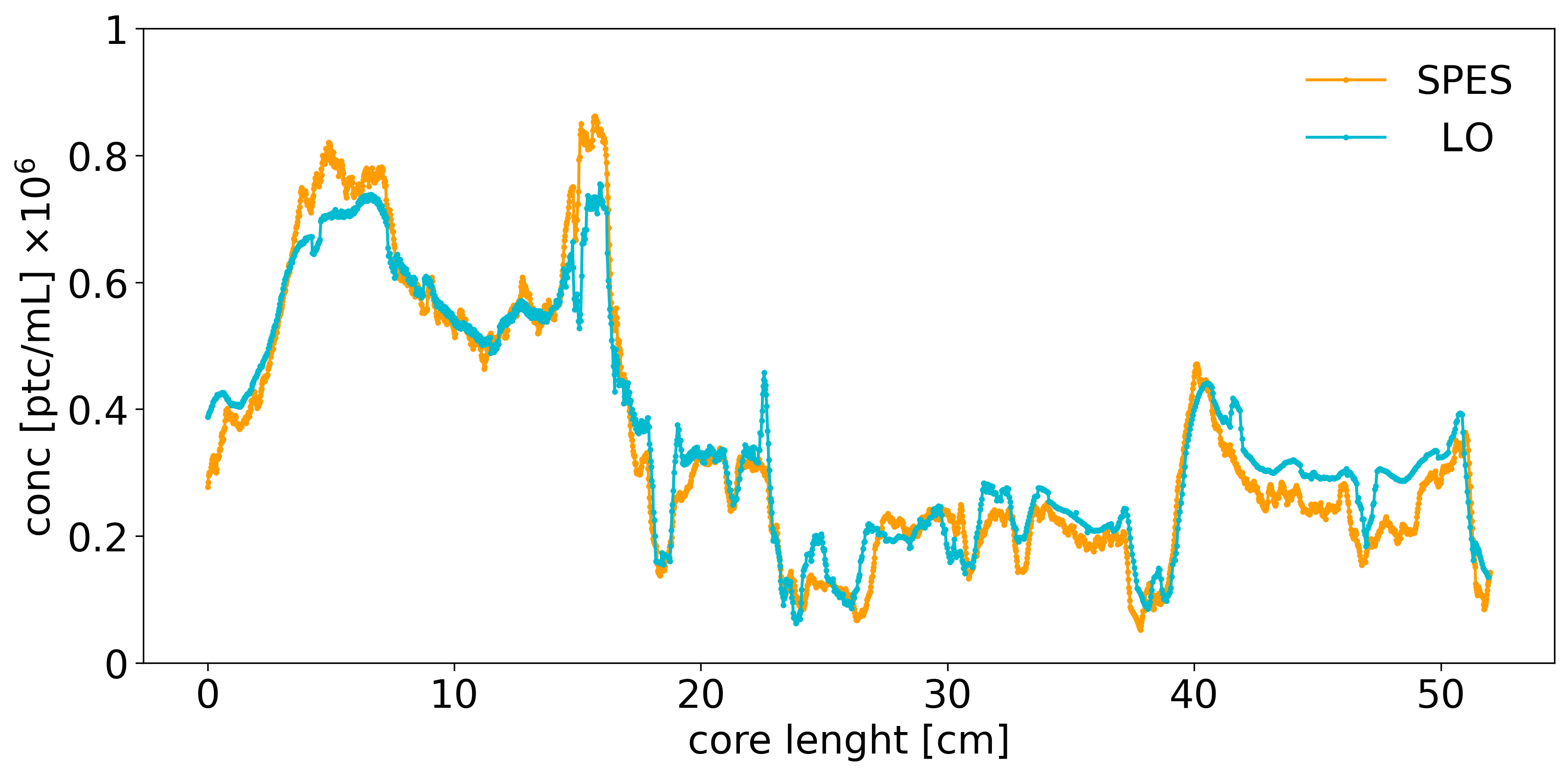} 
	\caption{(c) particle concentration profile}
	\end{subfigure}
	\caption{(a-b) PSDs measured by LO (red histogram, linear sampling) and SPES (blue histogram, logarithmic sampling) instruments before and after DEDALO application. Uncertainties, calculated as the square root of the number of particles measured per bin and expressed as 1 standard deviation, are reported as colored bands. (c) Sample particle concentrations, expressed in number of particles per mL of meltwater, as measured by the two devices.}
	\label{fig:rutor_example}
\end{figure}

As a practical application, we report the results obtained by applying DEDALO to perform accurate and high-resolution PSDs measurements in meltwater from the cryosphere, as commonly done through LO instruments \cite{lambert_1, lambert_2, wegner}. 
Figure \ref{fig:rutor_example} shows the results obtained from the CFA measurement of an ice core from the Rutor Glacier, Aosta Valley (Italy), approximately 60 cm long, under study with the scope of a characterization of the solid content of ice.
The PSD obtained from a traditional LO measurement differs significantly from SPES data (red histogram and blue histogram in Figure \ref{fig:rutor_example}a, respectively). By applying DEDALO's algorithm, this discrepancy can be compensated to a good extent (Figure \ref{fig:rutor_example}b). The refractive index set for this purpose is 1.45.
Quantitatively, in terms of Pearson's correlation coefficient, from an initial value of $r_{12}=0.56$ DEDALO allows for a peak of $r_{12}=0.82$.
Finally, Figure \ref{fig:rutor_example}c shows the particle concentration measured by both instruments as the depth of the ice core increases.
The close agreement between the two curves confirms the reliability of the LO measurement.

\section{Conclusions}
\label{sec:conclusion}

We have described a novel open-source Python-based GUI application that improves the procedures to recover the size of a particle from the data obtained through single-particle Light Obscuration sensors. 
By taking into account the most suitable refractive index based on the sample composition, we remove a strong assumption commonly adopted to extract size from the raw data in this kind of instruments. 
Since these data are not typically available, we use the calibration curve to recover the $C_\mathrm{ext}$ from the size values provided by the instrument.
Therefore, by introducing the sample refractive index, we are able to get a size that is reasonably much closer to the real one. As expected, the PSDs are appreciably affected by the choice of assuming one specific refractive index, as commonly exploited in commercial instruments. Comparing to the measurements obtained with an independent optical technique, we show that DEDALO is capable of partially compensating this effect.
Our method can improve the accuracy in particle sizing by Light Obscuration measurements in several fields \cite{LO_applcation1, LO_applcation2}. Our data clearly show that the PSDs obtained with DEDALO are more reliable compared to those compliant with the corresponding ISO standard \cite{ISO}.

Upon testing with different kinds of oil emulsions, DEDALO proved to be reliable and consistent over a wide range of refractive indices, from 1.46 up to 1.64, both for single refractive index solutions and in the case of mixtures composed of multiple substances.
We observed that some limitations arise for large values of the refractive index; this is mainly due to the oscillations in the $C_\mathrm{ext}$ trend as size increases, which become more and more frequent with increasing $n$ and hence alter the reconstruction.
In the specific case of mineral dust, for which the (effective) refractive index value can span from about $n=1.45$ to $n=1.52$, the effectiveness of DEDALO has been tested as well by analyzing a 60 cm ice core from the Rutor Glacier, Aosta Valley. By applying the proposed inversion algorithm, the correlation coefficient between the PSDs measured by the LO instrument and by another independent optical method (SPES) has more than doubled. In addition, the strong agreement of the particle concentration curves measured with the two techniques further confirms DEDALO reliability and high-resolution for PSDs measurements in meltwater from the cryosphere.
Even more generally, after the extensive validation of our code we believe that DEDALO could potentially have many further applications in pharmaceutical and industrial quality control processes, determining the size distribution of different types of organic samples: commercial plastic polymers such as PVC or PMMA ($n=1.5388$ and $n=1.4995$ at 670 nm respectively), some organic compounds including urea ($n=1.4873$) and cellulose ($n=1.4671$) and many hydrocarbons with refractive indices ranging from 1.32 up to 1.43 \cite{ref_index}.

DEDALO has been implemented with an ad hoc graphical user interface which easily allows to drive an in-line instrument. This application was initially developed with the aim of conducting reliable, high-resolution measurements of mineral dust stored in both the polar and Alpine cryosphere, within a collaboration with the paleoclimate and glaciology group at the University of Milano-Bicocca.

The algorithm is publicly available on the website of the Instrumental Optics Laboratory of the Physics Department of the University of Milan (\url{https://instrumentaloptics.fisica.unimi.it/dedalo/}) and on the GitHub page of the corresponding author (\url{https://github.com/LucaTeruzzi/DEDALO}). 
Being written in Python 3.10, the algorithm can run on all operative systems without compatibility issues. For Windows systems, DEDALO executable that does not need to be compiled manually is also available.
It will be continuously updated with new functionalities and cross-platform compatibility.


\appendix
\section{Oil refractive index dependence on wavelength}
\label{appendix_A}

For each oil considered in this work, the refractive index as a function of the wavelength ($\lambda$) has been computed through Cauchy equation

\begin{equation}
    n_\mathrm{oil}(\lambda) = A + B\dfrac{1}{\lambda^2} + C\dfrac{1}{\lambda^4} \quad ,
    \label{eq:cauchy_eqn}
\end{equation}

where the coefficients $A$, $B$ and $C$ are summarised in Table \ref{tab:cauhy_coefficients} \cite{cargille}:

\begin{table}[htbp]
\renewcommand{\arraystretch}{1.1}
\centering
\caption{Cauchy coefficients for each oil nominal refractive index.\label{tab:cauhy_coefficients}}
\smallskip
\begin{tabular}{c|c|c|c}
\hline\hline
\hspace{0.6cm} \textbf{Reference $n$} \hspace{0.6cm} & \hspace{0.6cm} \textbf{$A$} \hspace{0.6cm} & \hspace{0.6cm} \textbf{$B$ [\SI{}{\micro\meter^2}]} \hspace{0.6cm} & \hspace{0.6cm} \textbf{$C$ [\SI{}{\micro\meter^4}]} \hspace{0.6cm} \\
\hline\hline
1.46 & 1.45 & 4.07$\cdot$10$^3$ & 4.16$\cdot$10$^7$ \\
1.47 & 1.46 & 4.40$\cdot$10$^3$ & 6.54$\cdot$10$^7$ \\
1.48 & 1.47 & 4.73$\cdot$10$^3$ & 8.92$\cdot$10$^7$ \\
1.50 & 1.48 & 5.39$\cdot$10$^3$ & 1.37$\cdot$10$^8$ \\
1.51 & 1.49 & 5.72$\cdot$10$^3$ & 1.60$\cdot$10$^8$ \\
1.52 & 1.50 & 6.05$\cdot$10$^3$ & 1.84$\cdot$10$^8$ \\
1.53 & 1.51 & 6.37$\cdot$10$^3$ & 2.08$\cdot$10$^8$ \\
1.56 & 1.54 & 7.36$\cdot$10$^3$ & 2.79$\cdot$10$^8$ \\
1.58 & 1.55 & 8.05$\cdot$10$^3$ & 3.66$\cdot$10$^8$ \\
1.64 & 1.60 & 1.03$\cdot$10$^4$ & 8.11$\cdot$10$^8$ \\
\hline\hline
\end{tabular}
\end{table}

\begin{figure}[htbp]
	\centering
	\includegraphics[scale=0.55]{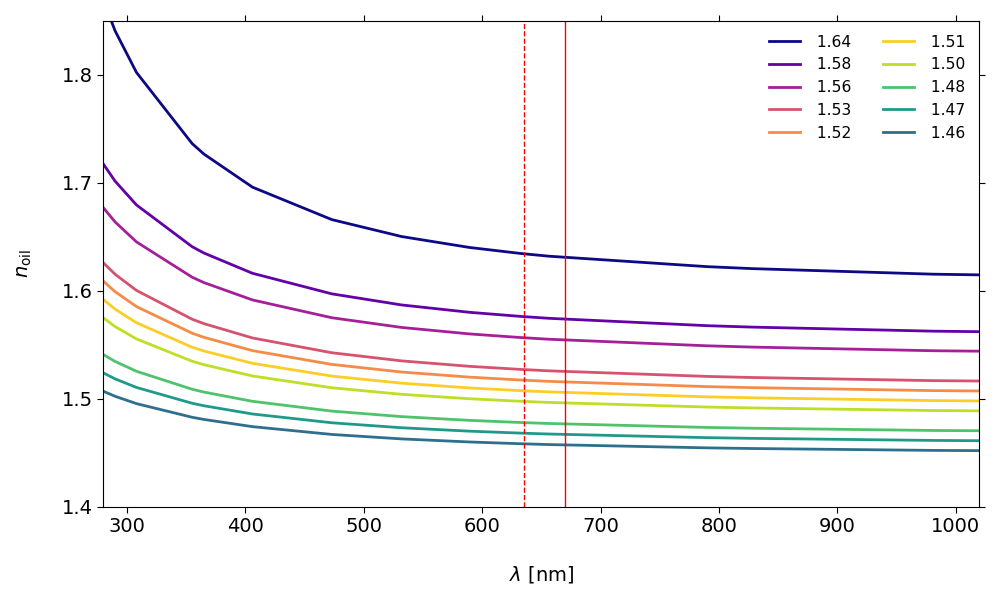} 
	\caption{Refractive index as a function of the wavelength for all the oil samples considered in this work, from $n=$1.46 (bottom-most curve, dark green) to $n=$1.64 (top-most curve, purple). The red solid line identifies the operative wavelength of the LO instrument (670 nm), while the red dashed one defines the SPES wavelength (635 nm). }
	\label{fig:oil_ref_index}
\end{figure}

Figure \ref{fig:oil_ref_index} shows the refractive index variation with wavelength for all the analyzed samples.
The two wavelengths corresponding to the operative values of LO ($\lambda_1=$ 670 nm) and SPES ($\lambda_2=$ 635 nm) instruments are also highlighted with a red solid line and a red dashed line respectively.
\newpage

In addition, in Table \ref{tab:n_vs_lambda} the refractive index values evaluated at $\lambda_1=$ 670 nm and $\lambda_2=$ 635 nm and their difference ($\Delta n_{12}$) are reported.

\begin{table}[htbp]
\renewcommand{\arraystretch}{1.1}
\centering
\caption{Comparison between the oil refractive index evaluated at the SPES and LO operative wavelengths (635 nm and 670 nm respectively).\label{tab:n_vs_lambda}}
\smallskip
\begin{tabular}{c|c|c|c}
\hline\hline
\hspace{0.6cm} \textbf{Reference $n$} \hspace{0.6cm} & \hspace{0.6cm} \textbf{635 nm} \hspace{0.6cm} & \hspace{0.6cm} \textbf{670 nm} \hspace{0.6cm} & \hspace{0.6cm} \textbf{$\Delta n_{12}$} \hspace{0.6cm} \\
\hline\hline
1.46 & 1.458 & 1.457 & 1$\cdot$10$^{-3}$\\
1.47 & 1.468 & 1.467 & 1$\cdot$10$^{-3}$\\
1.48 & 1.478 & 1.476 & 2$\cdot$10$^{-3}$\\
1.50 & 1.497 & 1.496 & 3$\cdot$10$^{-3}$\\
1.51 & 1.507 & 1.505 & 2$\cdot$10$^{-3}$\\
1.52 & 1.517 & 1.515 & 2$\cdot$10$^{-3}$\\
1.53 & 1.527 & 1.525 & 2$\cdot$10$^{-3}$\\
1.56 & 1.556 & 1.554 & 2$\cdot$10$^{-3}$\\
1.58 & 1.576 & 1.574 & 3$\cdot$10$^{-3}$\\
1.64 & 1.634 & 1.631 & 3$\cdot$10$^{-3}$\\
\hline\hline
\end{tabular}
\end{table}

The expected difference in refractive index, in the order of 10$^{-3}$, is clearly much smaller than the 
error derived by fitting the SPES data ($\sigma_n \sim$10$^{-2}$, as reported in Section \ref{sec:data}).
Thus, it is negligible to the correction to size inversion and can be disregarded to our purposes.


\acknowledgments

This work has is the result of a collaboration with the EuroCold Lab (European Cold Laboratory Facilities) and the paleoclimate and glaciology group at the University of Milan-Bicocca. We acknowledge Tiziano Sanvito and Marco Pallavera of EOS S.r.l. for their valuable support. 




\end{document}